\documentclass[twocolumn,superscriptaddress]{revtex4}
\usepackage{graphicx}
\usepackage{times}
\usepackage{amsmath}
\usepackage{amssymb}
\usepackage{units}
\usepackage{stmaryrd} 
\usepackage[compact]{titlesec}
\titlespacing{\section}{0pt}{*0}{*0}
\titlespacing{\subsection}{0pt}{*0}{*0}
\titlespacing{\subsubsection}{0pt}{*0}{*0}

\begin{document}
\preprint{0}

\title{Micro-metric electronic patterning of a topological band structure using a photon beam}

\author{E. Frantzeskakis}
\email{e.frantzeskakis@uva.nl} 
 \address{Van der Waals-Zeeman Institute, Institute of Physics (IoP), University of Amsterdam, Science Park 904, 1098 XH, Amsterdam, The Netherlands}

\author{N. de Jong}
\address{Van der Waals-Zeeman Institute, Institute of Physics (IoP), University of Amsterdam, Science Park 904, 1098 XH, Amsterdam, The Netherlands}

\author{B. Zwartsenberg}
\address{Van der Waals-Zeeman Institute, Institute of Physics (IoP), University of Amsterdam, Science Park 904, 1098 XH, Amsterdam, The Netherlands}

\author{Y. K. Huang}
\address{Van der Waals-Zeeman Institute, Institute of Physics (IoP), University of Amsterdam, Science Park 904, 1098 XH, Amsterdam, The Netherlands}

\author{T. V. Bay}
\address{Van der Waals-Zeeman Institute, Institute of Physics (IoP), University of Amsterdam, Science Park 904, 1098 XH, Amsterdam, The Netherlands}

\author{P. Pronk}
\address{Van der Waals-Zeeman Institute, Institute of Physics (IoP), University of Amsterdam, Science Park 904, 1098 XH, Amsterdam, The Netherlands}

\author{E. van Heumen}
\address{Van der Waals-Zeeman Institute, Institute of Physics (IoP), University of Amsterdam, Science Park 904, 1098 XH, Amsterdam, The Netherlands}

\author{D. Wu}
\address{Van der Waals-Zeeman Institute, Institute of Physics (IoP), University of Amsterdam, Science Park 904, 1098 XH, Amsterdam, The Netherlands}

\author{Y. Pan}
\address{Van der Waals-Zeeman Institute, Institute of Physics (IoP), University of Amsterdam, Science Park 904, 1098 XH, Amsterdam, The Netherlands}

\author{M. Radovic}
\address{Swiss Light Source, Paul Scherrer Institut, CH-5232 Villigen, Switzerland}
\address{SwissFEL, Paul Scherrer Institut, CH-5232 Villigen, Switzerland}

\author{N. C. Plumb}
\address{Swiss Light Source, Paul Scherrer Institut, CH-5232 Villigen, Switzerland}

\author{N. Xu}
\address{Swiss Light Source, Paul Scherrer Institut, CH-5232 Villigen, Switzerland}

\author{M. Shi}
\address{Swiss Light Source, Paul Scherrer Institut, CH-5232 Villigen, Switzerland}

\author{A. de Visser}
\address{Van der Waals-Zeeman Institute, Institute of Physics (IoP), University of Amsterdam, Science Park 904, 1098 XH, Amsterdam, The Netherlands}

\author{M. S. Golden}
\email{m.s.golden@uva.nl} 
\address{Van der Waals-Zeeman Institute, Institute of Physics (IoP), University of Amsterdam, Science Park 904, 1098 XH, Amsterdam, The Netherlands}

\date{\today}
\begin{abstract}
 In an ideal 3D topological insulator (TI), the bulk is insulating and the surface conducting due to the existence of metallic states that are localized on the surface; these are the topological surface states. Quaternary Bi-based compounds of Bi$_{2-\textmd{x}}$Sb$_{\textmd{x}}$Te$_{3-\textmd{y}}$Se$_{\textmd{y}}$ with finely-tuned bulk stoichiometries are good candidates for realizing ideal 3D TI behavior due to their bulk insulating character. However, despite its insulating bulk in transport experiments, the surface region of Bi$_{2-\textmd{x}}$Sb$_{\textmd{x}}$Te$_{3-\textmd{y}}$Se$_{\textmd{y}}$ crystals cleaved in ultrahigh vacuum also exhibits occupied states originating from the bulk conduction band. This is due to adsorbate-induced downward band-bending, a phenomenon known from other Bi-based 3D TIs.
Here we show, using angle-resolved photoemission, how an EUV light beam of moderate flux can be used to exclude these topologically trivial states from the Fermi level of Bi$_{1.46}$Sb$_{0.54}$Te$_{1.7}$Se$_{1.3}$ single crystals, thereby re-establishing the purely topological character of the low lying electronic states of the system. 
We furthermore prove that this process is highly local in nature in this bulk-insulating TI, and are thus able to imprint structures in the spatial energy landscape at the surface. We illustrate this by `writing' micron-sized letters in the Dirac point energy of the system.
\end{abstract}

\maketitle
\begin{widetext}

\section*{Introduction}

Topological insulators (TIs) are a novel state of quantum matter, serving as a platform for the observation of fundamental physics phenomena and possessing high potential for applications ranging from spintronics to quantum computation \cite{Hasan2010,Fu2007_2,Stern2013,Pesin2012}.
Their remarkable properties are linked to two-dimensional electronic states termed topological surface states (TSS) \cite{Xia2009,Zhang2009,Chen2009}, which are protected from backscattering and Anderson localization by non-magnetic impurities due to the combined effect of a helical spin texture and time-reversal invariance \cite{Roushan2009,Fu2007,Hsieh2009}.
In an applications context, controlled tunability of the electronic bands between regimes dominated by either the TSS or bulk-derived states of trivial topology is important.
Different approaches have been reported on single crystals of Bi-based TIs mainly involving surface decoration \cite{Hsieh2009,Chen2010,Zhu2011,King2011,Bahramy2012,Valla2012} and tuning the bulk stoichiometry \cite{Arakane2012,Ren2011}, both of which involve modification of the (surface) band structure across the whole sample.

Here we report a new approach on single crystals of TIs. Our method not only provides full control over the band structure at the surface of topological insulators but does so on a local, micron-scale level.
Our method requires (a) a bulk-insulating TI materials platform, and (b) a patterning tool - in our case an extreme-ultraviolet (EUV) photon beam of flux $>$10$^{21}$ photons$/(\textmd{s}$ $\textmd{m}^{2})$.
We use angle-resolved photoelectron spectroscopy (ARPES) to track deliberate modifications of the electronic structure of a quaternary TI compound
that possesses high bulk resistivity, namely Bi$_{1.46}$Sb$_{0.54}$Te$_{1.7}$Se$_{1.3}$ (BSTS1.46). In specific, we demonstrated successful imprinting of a pre-defined, micro-metric spatial pattern - the letters IoP - into the surface electronic band structure of the sample.  

Bi$_{2-\textmd{x}}$Sb$_{\textmd{x}}$Te$_{3-\textmd{y}}$Se$_{\textmd{y}}$ (BSTS) systems are excellent bulk insulators \cite{Ren2011,Pan2014}) and possess TSS, which display the characteristic dispersion relation called a Dirac cone \cite{Arakane2012,Neupane2012}. Transport experiments on single crystalline nanoflakes from stoichiometric and non-stoichiometric BSTS have proven its high potential for topological transport devices through the observation of surface dominated conduction \cite{Xia2013,Xu2014}, anisotropic magnetoresistance \cite{Sulaev2015}, and half-integer quantum Hall effect \cite{Xu2014}. Although such transport phenomena suggest that the topological surface state dominates the near-E$_{\textmd{F}}$ electronic band structure, we will show that the direct experimental view on the band dispersion of BSTS reveals a more complicated picture.
Our results for BSTS1.46 - i.e. the most insulating BSTS composition \cite{Pan2014} - are presented in Fig. 1. It is clear that surface decoration due to the adsorption of residual gas atoms from UHV (panels a-c) is the cause of downward band bending \cite{Zhang2012,Jong2015}. 
Thus, in practice, despite the bulk insulating nature of the crystal, within a near-surface region of thickness a few 10's of nm, the conduction band states are shifted below the Fermi energy (E$_{\textmd{F}}$), where they evolve into a parabolic band (Fig. 1c). This parabolic band is a signature of electrons confined in the potential-well defined by the surface potential and the band bending profile into the bulk.
Consequently, the real-life situation for a cleaved surface of BSTS (even in UHV), is that we have an insulating bulk, topologically-protected surface states and topologically-trivial surface-related states related to the dipping down in energy of the conduction band due to band bending.\\

\begin{figure}
  \centering
  \includegraphics[width = 10.5 cm]{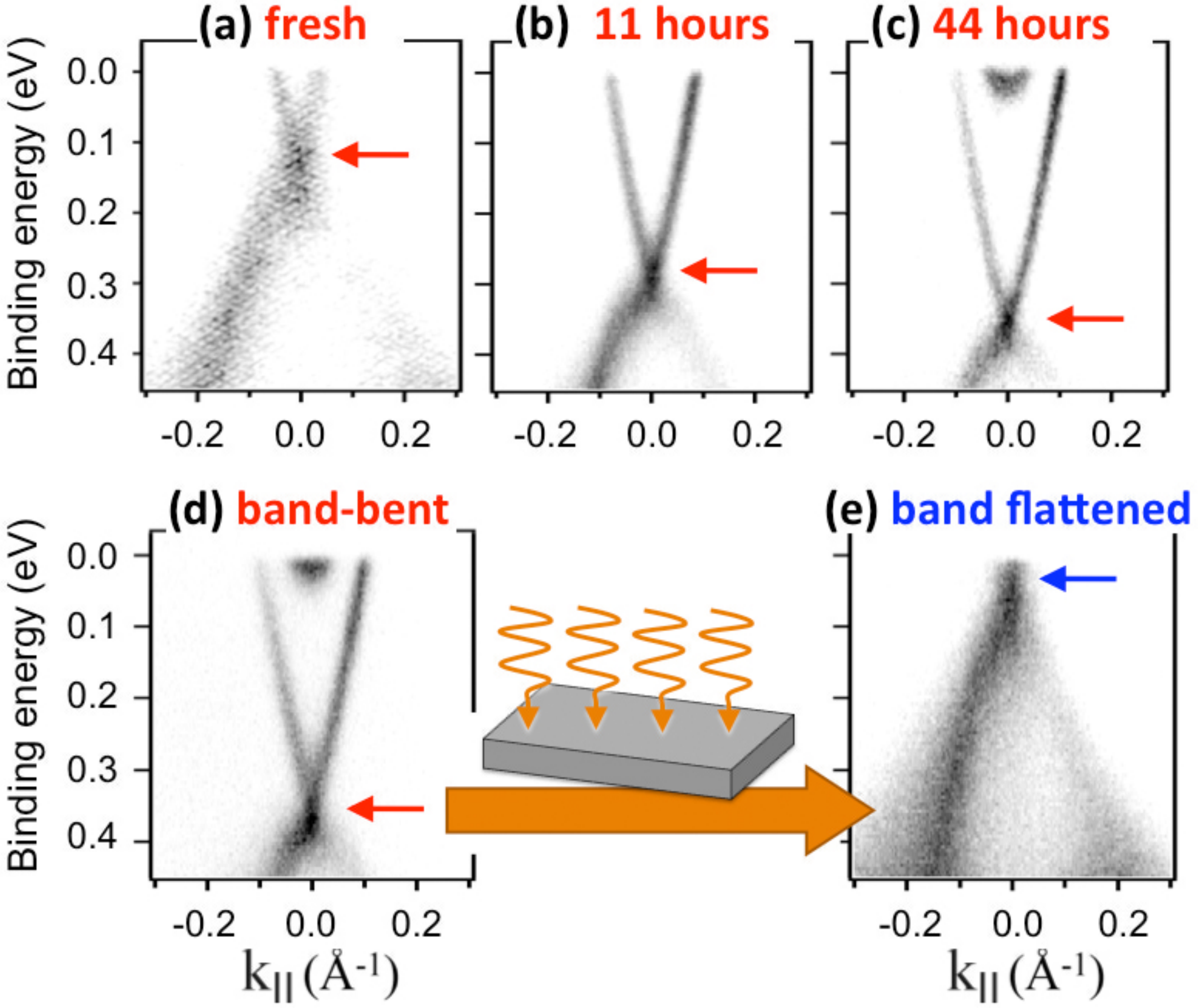}
  \caption{\textbf{Tunability of the electronic structure of bulk-insulating Bi$_{1.46}$Sb$_{0.54}$Te$_{1.7}$Se$_{1.3}$ by adatom adsorption and by exposure to high-flux photons.} (a-c) Near-E$_{\textmd{F}}$ electronic structure on increasing exposure to residual gases in a UHV environment.
(d,e) Strong changes occur in the near-E$_{\textmd{F}}$ electronic structure after long exposure to a high-fluence photon beam [flux 3.2 $\times$ 10$^{21}$ photons/(s m$^{2}$) for 4 hours]. After exposure, the Dirac point has shifted upward to lie very close to E$_{\textmd{F}}$, as indicated by the red/blue arrows. The sample temperature for all data was 17K.}\label{Fig1}
\end{figure}  

\section*{Results}

We now demonstrate that this dual character offers the chance to act on the electronic band structure at the micron spatial scale, with lifetimes exceeding five hours.
With band-bent BSTS as the starting point, we are able to selectively erase the topologically trivial surface states at the Fermi level, making use only of a beam of photons with energy exceeding the band gap of BSTS.
This makes BSTS's combination of a highly-insulating bulk and a metallic surface of mixed topological character (i.e. topologically trivial and non-trivial states) ideal for EUV electronic patterning. 

 \begin{figure*}
  \centering
  \includegraphics[width =17.8 cm]{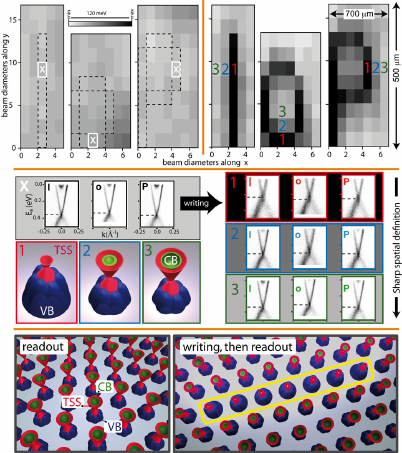}
  \caption{\textbf{Demonstration of micro-metric electronic patterning of the electronic states of the topological insulator Bi$_{1.46}$Sb$_{0.54}$Te$_{1.7}$Se$_{1.3}$
(BSTS1.46).} The three top-left panels represent the spatial dependence of the Dirac point energy (E$_{\textmd{D}}$) under
saturated band bending conditions, before high-fluence photon exposure.
The grey-scale shows that the spatial variations in this E$_{\textmd{D}}$-landscape are negligible.
Each pixel corresponds to an I(E,$k$)-measurement, with examples (white `X' locations) shown in the center-left, black-framed panel.  
The regions defined by the dashed lines were then exposed to a high-fluence photon beam. After low-fluence ARPES `readout', the new spatial E$_{\textmd{D}}$-landscape is shown in the top-right panels, indicating that the letters I, o and P (IoP, the acronym for Institute of Physics) have been successfully imprinted into the electronic states of the BSTS1.46 surface.
The numbers `1', `2', `3' on the post-writing areas label pixels on the written line, the nearest-neighbor and next-nearest neighbour sites, respectively, and the respective I(E,$k$) images are shown in the mid-right red-, blue- and green-framed boxes.
Comparison of the black- and red-framed I(E,$k$) images reveals that the writing process has pushed the topologically trivial band-bent states above E$_{\textmd{F}}$.
The neighboring pixels (blue-framed data-box) are only moderately affected, while the 2$^{\textmd{nd}}$-nearest neighbors (green-framed data-box) show
negligible change in E$_{\textmd{D}}$. These electronic band structures are summarized in a graphical schematic in the appropriate framed box. The lower two panels illustrate the readout of the band-bent starting situation and the situation after reading out the result of the writing process. Pre-exposure and post-exposure pixel images are made using the same grey-scale code (top-left grey-scale bar). 
The grey-scale pixel images in the upper panels are scaled in units of the beam diameter along the x(y) axes, which was 100(30) $\mu$m. 
The sample temperature was 17K and the photon flux was greater than 10$^{21}$ photons$/(\textmd{s}$ $\textmd{m}^{2})$. The fluence per pixel in the spatial E$_{\textmd{D}}$ maps used for readout was 40$\times$ lower than the corresponding fluence used for writing.}
\label{Fig2}
\end{figure*}

Comparison of Figs. 1d and 1e show the effect of a beam of photons with energy exceeding the band gap of BSTS and flux $>$10$^{21}$ photons/(s m$^{2}$).
The starting point is band bent BSTS1.46 (Fig. 1d) after a few hours in UHV at low temperature, with only minimal exposure to synchrotron radiation.
We noticed - during ARPES experiments that involved exposing the same surface to a photon flux of 3.2 $\times$ 10$^{21}$ photons/(s m$^{2}$) for four hours - that the band structure was altered, giving the data Fig. 1e. A pronounced energy shift of the TSS to lower binding energies is evident, whereby the Dirac point (blue arrow), now lies at an energy comparable to E$_{\textmd{F}}$.
Such photon-induced modifications of the downward band-bent electronic structure of TIs have been theoretically suggested \cite{Koleini2013} and experimentally observed by ARPES \cite{Kordyuk2011,Frantzeskakis2015,Yu2011}. They originate from the interplay of two effects: the photon-stimulated desorption of adsorbed atoms which were at the origin of band bending and the surface photovoltage effect \cite{Frantzeskakis2015}. These processes possess differing time-scales (see SI-2), but contribute together to the restoration of the charge distribution before the evolution of the adsorbate-induced band bending, by the removal of foreign electron donors from the sample surface (desorption), and the creation of electron hole pairs that separate along the $z$-direction thereby counteracting the adsorbate-induced potential (surface photovoltage) \cite{Kordyuk2011,Kronik1999,Alonso1989,Leibovitch1994}. 

The key message is that EUV illumination enables the tuning of the energy position of the electronic features so as to approach a situation where the surface band bending is negligible and the surface chemical potential is identical to the bulk (i.e. flat-band conditions). Our experiments show that the total energy shift tracks the total photon fluence until saturation is attained (see Fig. S1 in SI-1 and Ref. \cite{Frantzeskakis2015}).
This phenomenology allows us to carry out a quick (i.e. low photon fluence) I(E,$k$) measurement to `readout' the energy of the surface bands, without significant photon-induced effects, as has been done in Figs 1a-c. The photon fluence can be controlled by varying either the integrated exposure time or the flux itself. At intermediate exposure to high-flux illumination, significant spectral broadening accompanies the energy shift, as shown in SI-1. This broadening is absent in the initial and final stages of photon-induced band flattening (Fig. 1e and SI-1), and is due to the effects of inhomogeneous illumination within the beam profile.
The broadening can be successfully modelled (SI-2), and is an expression of the local nature of photon-induced changes, an observation which is at the core of the present study.
It is important to note that while photon-induced changes on BSTS are substantial and very local in character, they are less pronounced on Bi$_2$Se$_3$ and without any indication of sharp spatial definition (SI-3). The focus of what follows is on the spatial sensitivity of this process in BSTS, as this enables micro-metric electronic patterning of the band structure, as we will describe below.

Fig. 2 summarizes the key result of our research for a crystal of BSTS1.46. 
In the upper row of images, each rectangular pixel represents an I(E,$k$) image, recorded using an EUV beamspot of 100 $\mu$m (h) $\times$ 30 $\mu$m (v).
Representative I(E,$k$) images recorded from the individual pixels marked with a white `X' are shown, center-left, and display the characteristic linear TSS dispersion and signatures of the bulk valence and conduction bands.
Due to prior exposure to UHV, essentially saturated band-bent conditions are prevalent here, and the average binding energy of the Dirac point, E$_{\textmd{D}}$, is $\sim$370 meV.
The three grey-scale composite images in the top-left panel represent a spatial readout of E$_{\textmd{D}}$ (with 10 seconds measuring time per spatial pixel), and prove that there is a low spatial variation of E$_{\textmd{D}}$ in this `virgin' state over a mm scale area of the cleaved crystal surface.  
 
We subsequently expose pre-selected parts of the sample to a high-fluence photon beam.
The target areas are overlaid on the virgin state spatial readout using dashed lines and spell the letters ``I'' , ``o''  and ``P'', the acronym for Institute of Physics.
The fluence of the writing photon beam is 40$\times$ higher than the one used for E$_{\textmd{D}}$-readout. Comparison of the top-left and top-right spatial maps shows that after high fluence exposure, the selected structures have been successfully imprinted into the E$_{\textmd{D}}$ spatial landscape.

In the stack of data panels on the right-side of Fig. 2, the I(E,$k$) images from the three locations centered on the letters (marked with a red `1' and shown in the red-outlined black panel) make it clear that E$_{\textmd{D}}$ has shifted upward by some 150 meV and consequently the bulk conduction band is absent, now being above E$_{\textmd{F}}$.
After photon-induced writing, the nearest-neighbor (NN, blue `2') and next-nearest-neighbor (NNN, green `3') pixels are hardly affected, as can be seen in both the spatial maps and the corresponding I(E,$k$) images.
The NNN pixels show essentially no change with respect to the virgin state, and the NN pixels show only a very modest energy shift of order 20 meV. Each of the situations (target-, NN- and NNN-pixel) is sketched as an artist's impression underneath the I(E,$k$) images of the pristine state.
For both NN and NNN locations, besides the valence band, both the TSS and the surface-confined bulk conduction band states are still visible at binding energies $\le$E$_{\textmd{F}}$.  
So, what the data of Fig. 2 show is that for bulk insulating BSTS, the non-topological surface states can be removed along a spatial trajectory that can be chosen at will, leaving behind pure TSS's at E$_{\textmd{F}}$, which will dominate the transport properties of the system at these locations. 

The lowest panels of Fig. 2 show a schematic of the writing process. Bottom-left: the starting-point is a band bent, bulk-insulating TI, with high spatial homogeneity as regards its electronic structure. Bottom-right: application of a EUV beam on specific sample locations (within the yellow frame) shifts the bulk conduction band states (green) to the unoccupied part of the spectrum, leaving only the valence band states at higher binding energy and the TSS at E$_{\textmd{F}}$. 
Fig. 2 (and SI-2) makes it clear that the spatial extent of the photon-induced changes is controlled by the dimensions of the EUV light beam.    
This means that with the use of more finely focused beams or masks, true nanometric patterning of the electronic states should be within reach.   

Fig. 2 shows that this spatially-resolved tunability of the electronic band structure is a novel and additional `knob', giving us control over the topological states in single crystals of real bulk insulating TIs.

\begin{figure}[!b]
  \centering
  \includegraphics[width = 10.5 cm]{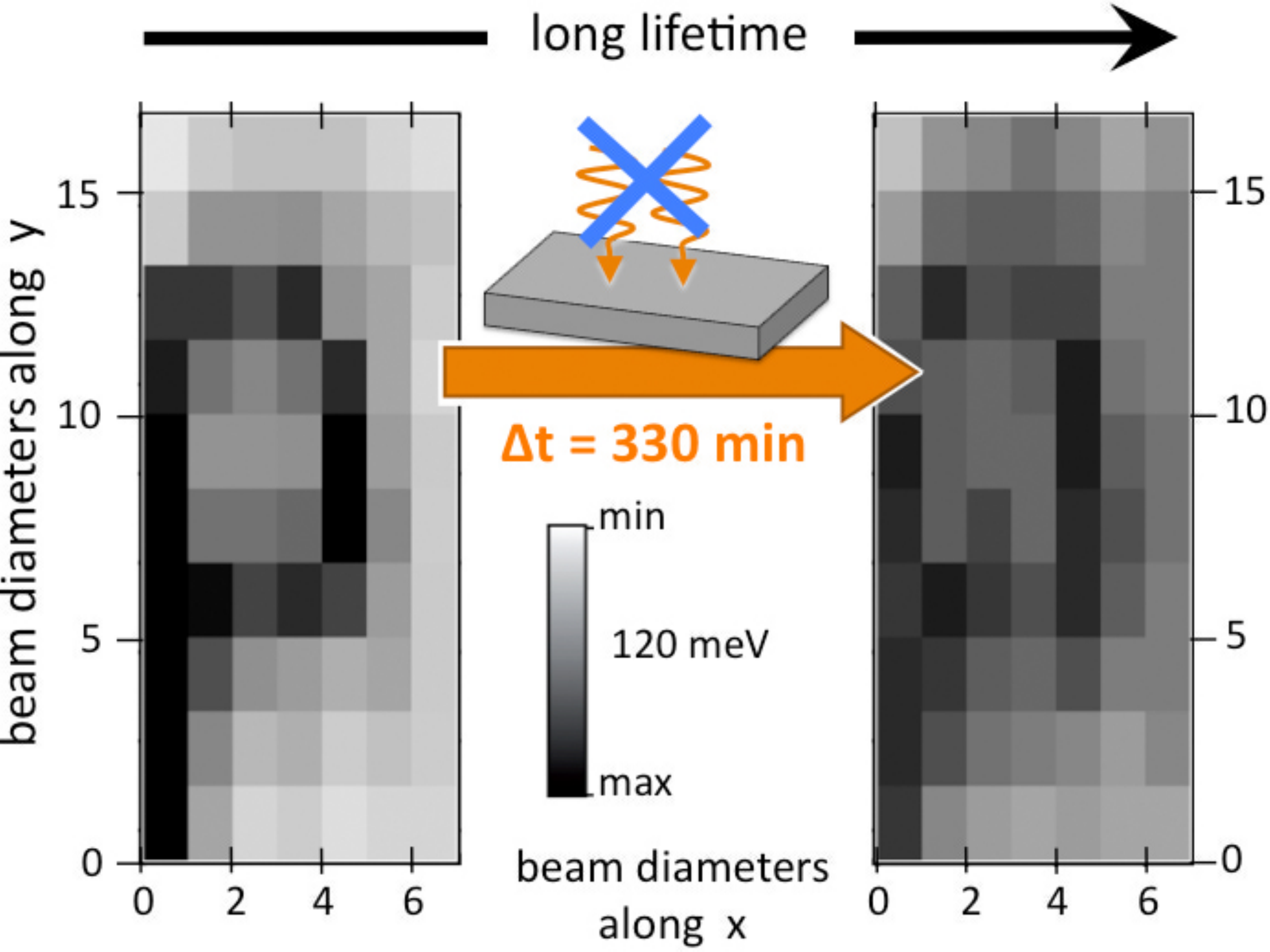}
  \caption{\textbf{The lifetime of photon-induced micro-metric patterns.} (left) The imprinted character is ``P'' reproduced from Fig. 2. The grey-scale encodes the energy shift of E$_{\textmd{D}}$ after exposure to an EUV beam of high photon fluence (see text). (right) Readout of the same area after leaving the sample in the dark for 6.5 hours. The sample temperature was 17K.}\label{Fig3}
\end{figure}  

The data shown in Fig. 3 address the (low temperature) lifetime of the photon-induced micron-sized structures. The left-hand panel shows the same `P' as was imaged in Fig. 2. 
This sample area then remained unexposed to external illumination until a second readout operation was carried out 330 min later, the result of which is shown in the right-hand panel of Fig. 3.
There is a reduction in the definition of the pattern, signaling finite in-plane diffusion of the photon-induced carriers and the slow reversibility of the writing process. Nevertheless, the imprinted pattern is still clearly visible at low temperatures after more than six hours without illumination.\\

\section*{Discussion}

In the following, we discuss the outlook for further improving this approach and touch on a few open issues.
Firstly, how far could the spatial resolution be improved? 
The data argue for a strong lateral confinement of the photon-induced charges in the bulk and near-surface region of BSTS at low temperature, as the written structures (Fig. 2) are as small as the photon beam used to write them.
The simulations discussed in SI-2 show that the system reacts to the differing photon exposures {\it within} the beam profile, pointing to possible sub-micron length scales upon further reduction of the beam size and control over the beam shape.
We mention that nano-ARPES measurements would be one way of providing a spectroscopic feasibility test \cite{Frantzeskakis2012,Bostwick2012} of future nanoscale patterning applications in TIs.

The second question is: can the lifetime of the photon-based manipulation of the electronic structure be extended?
The structures presented here were written in a serial, pixel-by-pixel process, during (and after) which, the sample surface is continuously exposed to residual gases, whose adsorption tends to increase, rather than flatten the downward band bending.
A parallel strategy in which different sample locations are exposed in a multi-beam device or one with a line focus, would significantly decrease the total time taken to imprint a pattern and would thus improve the `active' lifetime of the written pattern.
Periodic re-illumination or even permanent illumination could be considered to prolong the lifetime of the modification of the electronic structure indefinitely.

A final question: is a 3$^{\textmd{rd}}$ generation electron storage ring as a light source a necessary condition for this patterning approach? 
In our case, it was convenient to use the same photon source as both a patterning tool and (at reduced fluence) a readout tool.
However, one may use other photon sources capable of achieving comparable fluxes to those reported here. Interestingly a flux of 10$^{21}$ photons/(s m$^{2}$) is within the reach of standard laboratory lasers or LED sources. Such sources can easily satisfy the energy requirement of the underlying effects, namely excitation energies between a few hundreds meV and a few eV to desorb weakly bonded adatoms and super-band-gap (i.e. $>$ 400 meV) photons to induce surface photovoltage effects \cite{Kronik1999}.

Consequently, given the feasibility of photon-beam induced modification of the electronic structure at the surface of bulk-insulating TIs using readily available light sources, the way ahead looks clear for exploring how such manipulation of the electronic landscape can be exploited in transport and device-based experiments. Despite the high potential of nanoflakes from bulk-insulating TIs for topological transport devices \cite{Xia2013,Xu2014,Sulaev2015}, already some transport experiments have been interpreted as showing complications from topologically trivial, band-bending induced 2DEG states, besides the TSS \cite{Lee2012}.
Thus, even global spatial tuning of the surface chemical potential may prove to be of use. 
When considering adding the spatial structuring of the electronic states to transport experiments, careful experimental design will be required so as to avoid short-circuiting the illuminated `topological only' transport channel by the non-illuminated regions which maintain their metallic character and topologically trivial states at the Fermi level.
In transport experiments in which the active region is spatially defined - for example the gap between superconducting electrodes aimed at injecting a super-current through the TI's TSS \cite{Veldhorst2012J,Veldhorst2012E,Snelder2014} - the patterning technique introduced here could be used to selectively `wipe' the topologically trivial states from one junction. The superconductivity in the junction electrodes will re-establish itself very quickly compared to the lifetime of the writing effects (see Fig. 3) and then the illuminated SC-TSS-SC junction can be compared with the control SC-TSS+CB-SC junction created a few photon beam diameters away on the surface of the very same topological insulator crystal.      

In summary, the ARPES data presented here from the 3D, bulk insulating topological insulator Bi$_{1.46}$Sb$_{0.54}$Te$_{1.7}$Se$_{1.3}$ present unequivocal evidence that it is possible to exploit a beam of super band-gap photons (here EUV photons with flux exceeding 10$^{21}$ photons/[s m$^{2}$]) to reverse the downward band bending that results in the bulk conduction band states dipping below E$_{\textmd{F}}$ in the surface region. 
We show that such a photon beam can write micron-sized spatial patterns of arbitrary shape in the energy landscape of the topological electronic states in the surface region.
This electronic patterning persists on a timescale of several hours, and thus use of this photon-induced reversal of the occupation of surface-confined bulk conduction band states presents a novel route providing local control over the topological character of the states at the Fermi level of bulk-insulating TIs.\\

\section*{Methods}

The Bi$_{2-\textmd{x}}$Sb$_{\textmd{x}}$Te$_{3-\textmd{y}}$Se$_{\textmd{y}}$ crystals were grown in Amsterdam using the Bridgman technique. High purity elements were melted in evacuated, sealed quartz tubes at 850$^{\textmd{o}}$C and allowed to mix for 24 hours. The mix was subsequently allowed to cool down with a cooling rate of 3$^{\textmd{o}}$C per hour. Samples were cleaved at 17K and at a pressure better than 5$\times$10$^{-11}$ mbar. Temperature-dependent resistivity curves of Bi$_{1.46}$Sb$_{0.54}$Te$_{1.7}$Se$_{1.3}$ show an insulator-like upturn at low temperatures with a typical saturation value of 10 $\Omega$cm \cite{Pan2014,Frantzeskakis2015}.

ARPES experiments were conducted at the SIS beamline of the Swiss Light Source using a VG SCIENTA R4000 elecron analyzer. The photon energy was set to 27 eV for BSTS and 30 eV for Bi$_{2}$Se$_{3}$ (SI-3).
The photon flux was in the order of 10$^{21}$ photons$/(\textmd{s}$ $\textmd{m}^{2})$ while the total fluence could be modified at will by varying the total exposure time.
Similar experimental results were obtained by modifying the fluence via reduction of the flux itself (i.e. by a controlled detuning of the undulator harmonic energy \textit{vs.} the nominal monochromator energy), while keeping the exposure time constant. The temperature during measurements was 17K and the experimental energy resolution was set to 15 meV. The size of the beam spot was 100 $\mu$m $\times$ 30 $\mu$m. The samples were mounted on a cryo-manipulator (CARVING), offering three rotational degrees of freedom and excellent mechanical stability (i.e. negligible backlash), as was repeatedly verified using ARPES. The spatial resolution along the three axes of translation was 5 $\mu$m. An accurate energy position of the Fermi level was determined by evaporated Au thin films in contact with the sample holder.

\vspace{7 mm}

 \section*{Acknowledgements}
This work is part of the research program of the Foundation for Fundamental Research on Matter (FOM), which is part of the Netherlands Organization for Scientific Research (NWO). E.v.H. acknowledges support from the NWO Veni program. In addition, the research leading to these results has received funding from the European Community's Seventh Framework Programme (FP7/2007-2013) under grant agreement no. 312284 (CALIPSO).\\

 \section*{Author Contributions}
E.F., N.d.J. and M.S.G. conceived and planned the experiments. E.F., N.d.J., E.v.H., B.Z., T.V.B. and P.P. performed the ARPES experiments. Y.K.H. and D.W. grew the crystals. Y.P. and A.d.V. characterized the samples by transport measurements. M.R., N.C.P., N.X. and M.S. provided beamline support. E.F. performed the data analysis. B.Z. and E.F. did the simulations included in the supplementary information. E.F., E.v.H., N.d.J. and M.S.G. developed the interpretive framework for the data. E.F. and M.S.G. wrote the manuscript with discussions and comments from all co-authors.\\

 \section*{Additional Information}
 \subsection*{Competing Financial Interests}
 The authors declare no competing financial interests.

\end{widetext}

\end{document}